\documentclass[aps,prb,reprint,amsmath,superscriptaddress,longbibliography]{revtex4-2}
\usepackage{graphicx}
\usepackage{enumerate}
\usepackage{refcount}
\usepackage{fmtcount}
\usepackage{amssymb}
\usepackage{hyperref}
\usepackage{enumitem}

\usepackage[svgnames]{xcolor}
\usepackage[normalem]{ulem}
\usepackage{bm}

\hypersetup{pdfstartview=FitH,pdfpagemode=UseOutlines,colorlinks,
citecolor=blue,linkcolor=blue,urlcolor=blue}

\definecolor{blue(munsell)}{rgb}{0.0, 0.5, 0.69}

\def\sro{SrRu$_2$O$_4$}

\def\stf{$\kappa$-STF$_x$}
\def\STF{$\kappa$-[(BEDT-STF)$_x$(BEDT-TTF)$_{1-x}$]$\rm _2 Cu_2 (CN)_3$}

\def\Cu{$\kappa$-(BEDT-TTF)$_2$Cu$_2$(CN)$_3$}

\def\aI3{$\alpha$-I$_3$}

\begin{document}

\title{Universal relation between residual resistivity and $A$ coefficient in correlated metals}

\author{Anna Yu. Efimova}
\affiliation{Department of Quantum Matter Physics, University of Geneva, CH-1211 Geneva, Switzerland}
\author{Yohei Saito}
\affiliation{Department of Physics, Graduate School of Science, Hokkaido University, Sapporo 060-0810, Japan}
\author{Atsushi Kawamoto}
\affiliation{Department of Physics, Graduate School of Science, Hokkaido University, Sapporo 060-0810, Japan}
\author{Martin Dressel}
\affiliation{Universität Stuttgart, Pfaffenwaldring 57, D-70569 Stuttgart, Germany}
\author{Louk Rademaker*}
\affiliation{Department of Quantum Matter Physics, University of Geneva, CH-1211 Geneva, Switzerland}
\affiliation{Institute-Lorentz for Theoretical Physics, Leiden University, PO Box 9506, Leiden, NL-2300, The Netherlands}
\author{Andrej Pustogow}
\thanks{These authors contributed equally to this work. Correspondence: louk.rademaker@gmail.com; pustogow@ifp.tuwien.ac.at}
\affiliation{Institute of Solid State Physics, TU Wien, 1040 Vienna, Austria}

\date{\today}

\maketitle



{\bf The effects of strong electronic correlations and disorder are crucial for emergent phenomena such as unconventional superconductivity, metal–insulator transitions, and quantum criticality. While both are omnipresent in real materials, their individual impacts on charge transport remain elusive.
To disentangle their respective roles, we have independently varied the degree of randomness and the strength of electronic correlations -- by chemical substitution and physical pressure, respectively -- within the metallic phase nearby a Mott-insulating state. We find a distinct correlation dependence of the disorder‐dependent residual resistivity $\rho_0$ in the Fermi-liquid regime
\(
  \rho(T)=\rho_0 + A T^2,
\)
where $A\propto (m^{\star}/m)^2$ quantifies the electronic mass enhancement. Contrary to conventional expectations, we observe that at fixed disorder level $\rho_0$ grows {\em linearly} with $A$. This scaling can be understood in terms of chemical‐potential fluctuations with variance $\sigma_\mu^2$, yielding
\(
  \rho_0 \propto A\,\sigma_\mu^2.
\)
By comparing our findings to transport data on other organic Mott systems, oxides, heavy‐fermion compounds, and moiré materials, we demonstrate that this new relation between residual resistivity and mass enhancement is a universal feature of correlated metals.
}


The combined impact of electron-electron interactions and disorder on charge transport in correlated metals remains one of the most intensely studied, yet still unsolved issues of quantum matter physics. Of central importance is Landau's Fermi-liquid (FL) theory~\cite{Landau1956}, that explains the ubiquitous quadratic temperature dependence of the electrical resistivity in correlated metals, 
\begin{equation}
    \rho (T) = \rho_0 + A T^2.
    \label{Eq-resistivity}
\end{equation}
The commonly held wisdom is that $\rho_0$ reflects disorder scattering at $T \rightarrow 0$ whereas the $A$ coefficient incorporates all correlation dependence. In particular,  $A\propto (m^*/m)^2$ via the Kadowaki-Woods relation~\cite{Kadowaki1986,Jacko2009}, as the Sommerfeld coefficient of the specific heat $\gamma$ is proportional to the interaction-induced enhancement of quasiparticle (QP) mass $m^*/m$.

\begin{figure}[b!]
 \centering
 \includegraphics[width=0.95\columnwidth]{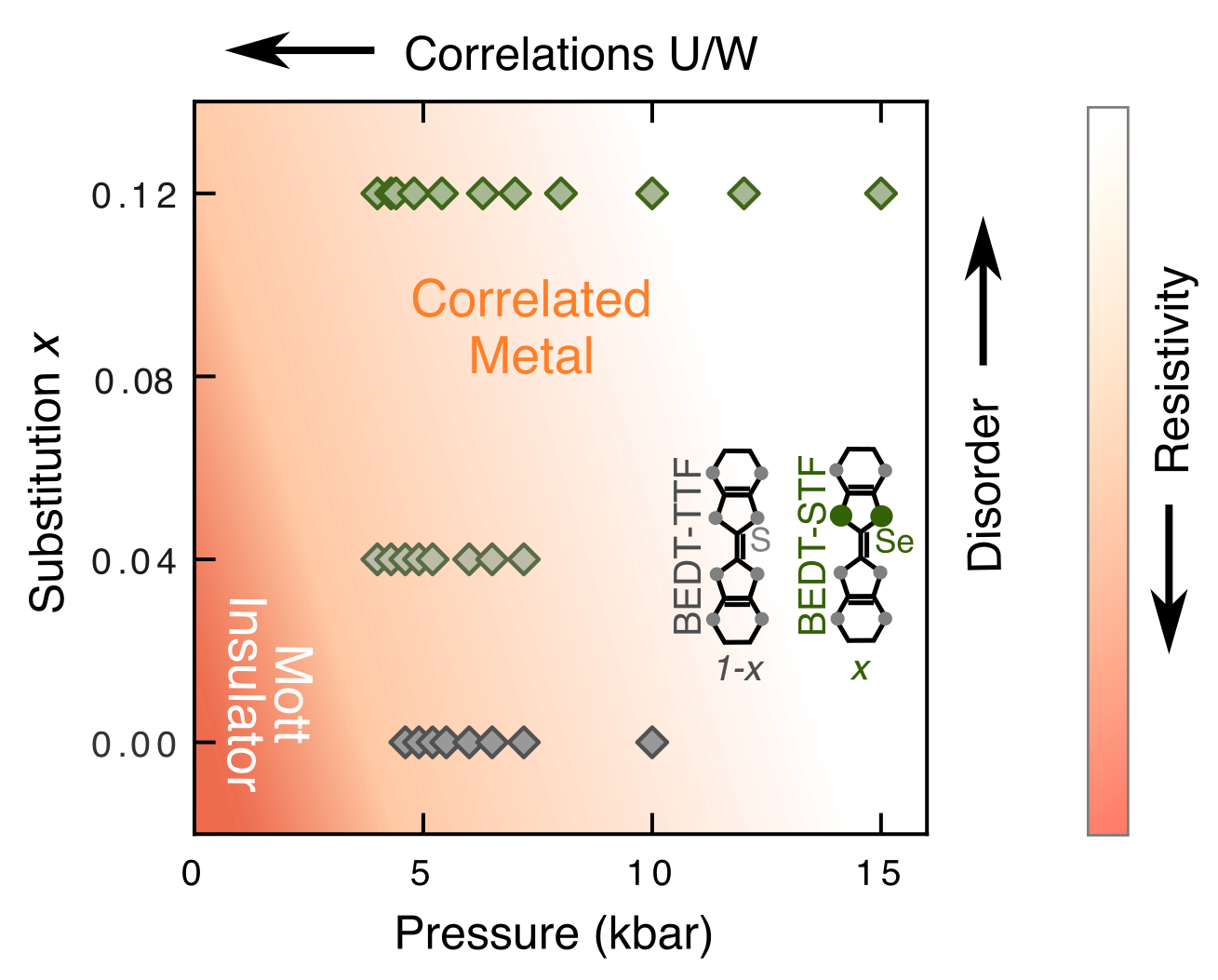}
 \caption{\textbf{Tuning of correlation strength and disorder in a Fermi liquid.} The symbols indicate the multi-dimensional phase space of the correlated metallic state nearby a Mott insulator~\cite{Pustogow2018,Pustogow2021-Landau,Pustogow2021-percolation} studied here in pressure $p$ (correlations $U/W$, hence $m^*/m$) and partial BEDT-STF substitution $x$ (disorder). Pressurizing single crystals ($x=0.00$, $0.04$, $0.12$) allows pure correlation tuning at different fixed disorder levels. 
 }
\label{intuition}
\end{figure}

\begin{figure*}
 \centering
 \includegraphics[width=2.1\columnwidth]{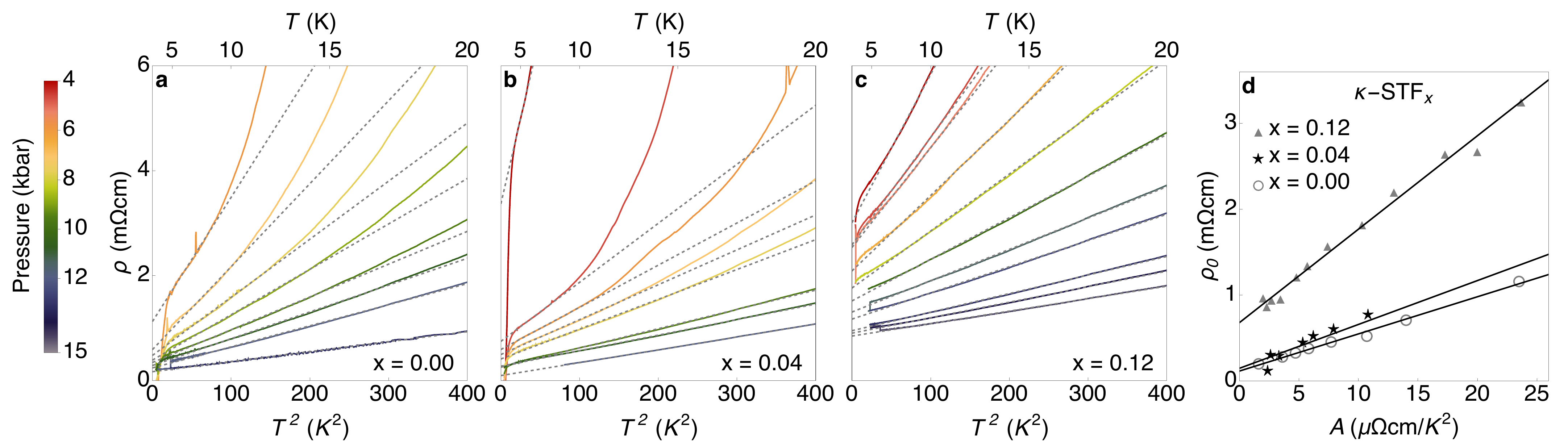}
 \caption{\textbf{Pressure-dependent dc resistivity of \STF.} \textbf{a-c} The plots of $\rho(T)$ versus $T^2$ for \stf\ with $x=$0.00, 0.04 and 0.12 reveal an increase of $A$ and also $\rho_0$ while approaching the metal-insulator transition (upon decreasing $p$). Dashed lines indicate fits to FL behavior (Eq.~\ref{Eq-resistivity}), from which both $A$ and $\rho_0$ were extracted. The pressures indicated here correspond to the $300$~K values~\cite{Rosslhuber2021,Pustogow2021-percolation}. \textbf{d} The residual resistivity $\rho_0$ scales linearly with the $A$-coefficient, with an offset $\rho_{00}$ that quantifies the sample-specific disorder scattering. The slope $d\rho_0/dA$ grows with increasing $x$.
 }
\label{STF-resistivity-all}
\end{figure*}


While deviations from the $T^2$ dependence in Eq.~\eqref{Eq-resistivity} in {\em bad} and {\em strange} metals have been vividly discussed over the past two decades~\cite{Gunnarsson2003,Hussey2004,Hartnoll2015,Phillips2023}, the residual resistivity has received less attention as it is supposed to be independent of electronic correlations~\cite{Coleman.2015p0w}. This can be seen in a standard Drude picture $\rho_0 = \frac{m}{ne^2\tau_0} = \frac{m^*}{ne^2\tau_0^*} $, as the enhancements of the effective mass $m^* = m/Z$ and the disorder scattering time $\tau_0^* = \tau_0/Z$ with diminishing QP weight $Z$ cancel each other. 
Opposite to that, here we report a pronounced increase of residual resistivity as correlations get more pronounced. We establish the linear relation $\rho_0\propto A$ in diverse correlated electron systems over many orders of magnitude and relate it to chemical potential fluctuations.

To experimentally investigate their impacts on $A$ and $\rho_0$, we have systematically tuned correlation strength and disorder {\em separately} in organic Mott systems, see Fig.~\ref{intuition}. In these versatile materials, disorder can be varied through chemical substitution~\cite{Kobayashi2016,Saito2018,Pustogow2021-Landau,Pustogow2021-percolation,Saito2021,Saito2021a,Yesil2023,Chen2025} or via x-ray irradiation ~\cite{Sasaki2012,Furukawa2015,Urai2020,*Urai2019,Yamamoto2020}.
On the other hand, electronic correlations are commonly tuned by external parameters such as hydrostatic pressure 
~\cite{Lefebvre2000,Limelette2003a,Kurosaki2005,Furukawa2015,Furukawa2018,Pustogow2018}, strain 
\cite{Hicks2014,Steppke2017,Luo2019,Chronister2022}, magnetic field 
~\cite{Custers2003,Lohneysen2007,Ronning.2005, Kushwaha2019} or displacement field.  
The paradigmatic compound \Cu\ serves as a model system for the pressure-tuned Mott metal-insulator transition~\cite{Kurosaki2005,Furukawa2015,Furukawa2018,Pustogow2018,Pustogow2021-percolation,Pustogow2023}, which hosts an extended FL phase at high pressure. 
Here we study the low-temperature resistivity in the correlated metallic phase of \STF\ (abbreviated \stf ). Note that, apart from changing the amount of disorder, partial substitution $x$ of the sulfur-based BEDT–TTF molecules by selenium-containing BEDT–STF donors also tunes the correlation strength $U/W$~\cite{Saito2018,Pustogow2021-Landau,Pustogow2021-percolation,Saito2021,Saito2021a,Yesil2023,Chen2025}. Crucially, upon pressure the increase of \(W\) reduces $U/W$ within a \stf\ single crystal without altering the impurity concentration; the on‐site Coulomb repulsion \(U\) remains essentially unchanged under compression~\cite{Pustogow2018,Saito2021,Pustogow2021-percolation,Dressel2020}. This way, by applying pressure we have selectively probed the correlation dependence of $\rho(T)$ at distinct disorder levels, covering the phase space indicated in 
Fig.~\ref{intuition}. 







Fig.~\ref{STF-resistivity-all}a-c shows our dc resistivity results for hydrostatic pressure applied in an oil pressure cell~\cite{Rosslhuber2021}. Superconductivity is found for all three samples~\cite{Saito2021a}. Increasing pressure suppresses the insulating and superconducting behavior, stabilizing metallic resistivity at low temperatures. Our analysis focuses on this metallic regime. Upon approaching low $T$, FL behavior with quadratic temperature dependence is clearly visible. 
Note that the $\rho\propto T^2$ regime extends approximately up to the Mott-Ioffe-Regel (MIR) limit $\rho_{\rm MIR} =hd/e^2= 3.8$~m$\Omega$cm~\cite{Hussey2004,Takenaka2005,Pustogow2021-Landau,Saito2021}. 

Upon fitting the low-temperature FL behavior in the $T^2$ plots of Fig.~\ref{STF-resistivity-all}a-c, we find that $A$ decreases with increasing $p$, in accordance to previous studies on \Cu\ under pressure and strain \cite{Kurosaki2005,Shimizu2011,Furukawa2018,Pustogow2021-Landau,Pustogow2021-percolation}. In addition, we extracted the residual resistivity $\rho_0$ according to Eq.~\ref{Eq-resistivity}, and find that it is systematically reduced with higher pressure. Strikingly, by plotting the two coefficients in Fig.~\ref{STF-resistivity-all}d we uncover a linear relationship $\rho_0 \propto A$, in defiance of the existing textbook theory for the residual resistivity~\cite{Coleman.2015p0w}. The dramatic increase of $\rho_0 (p)$ close to the Mott transition clearly indicates that it $does$ exhibit a correlation dependence.


In the limit $A\rightarrow 0$, the linear relation extrapolates to a 'residual' residual resistivity $\rho_{00}$. We associate it with the particular amount of disorder in each single crystal -- which remains unaffected by external pressure. The increase of $\rho_{00}$ with $x$ indicates that chemical substitution indeed adds randomness to the system independent of the interaction strength. 
Furthermore, we notice that also the slope $d\rho_0/dA$ increases with $x$, suggesting that it is correlated with disorder strength. 

As we elaborate in the following, the here observed linear relationship $\rho_0 \propto A$ results from the {\em frequency dependence} of the scattering rate $\tau^{-1}$. 
First, we go back to the fundamentals of FL theory~\cite{Landau1956,Gurzhi1959}, stating that the QP scattering rate grows quadratically with energy,
\begin{equation}
	\tau_{\rm FL}^{-1} (\omega, T) = \tau_{00}^{-1} + A' \left(\omega^2 + (\pi T)^2 \right) + \ldots,
	\label{Eq:FreqDepScattering}
\end{equation}
where $\tau_{00}$ is a frequency- and temperature-independent contribution from disorder scattering.
From Eq.~\ref{Eq:FreqDepScattering}, we extract the resistivity using the Kubo formula, \cite{Kubo1957,Fetter1971,Jacko2009,Fratini2023}, which reads for a parabolic band dispersion
\begin{equation}
	\rho (T) = \left[ \frac{ne^2}{m} \int d\omega \; \left(-\dfrac{\partial f(\omega)}{\partial\omega}\right) \; \tau(\omega,T) \right]^{-1},
	\label{Eq:Kubo}
\end{equation}
where $f(\omega)$ is the Fermi-Dirac distribution function, and $m$ the bare, unrenormalized mass of the electronic band. 
At low temperatures, the scattering rate of Eq.~\eqref{Eq:FreqDepScattering} yields a resistivity $\rho(T) = \rho_0 + AT^2$ with residual resistivity $\rho_0 = \rho_{00} \equiv \frac{m}{ne^2} \tau_{00}^{-1}$ and the coefficient $A = \frac{m}{ne^2} \frac{4 \pi^2}{3} A'$~\cite{SupplInfo}. Therefore, in this picture, the prefactor $A'$ of the $\omega^2$-term in the scattering rate is proportional to the $A$-coefficient of the resistivity, but the residual resistivity $\rho_0$ is independent of electron-electron interactions.

\begin{figure}
 \centering
 \includegraphics[width=0.5\textwidth]{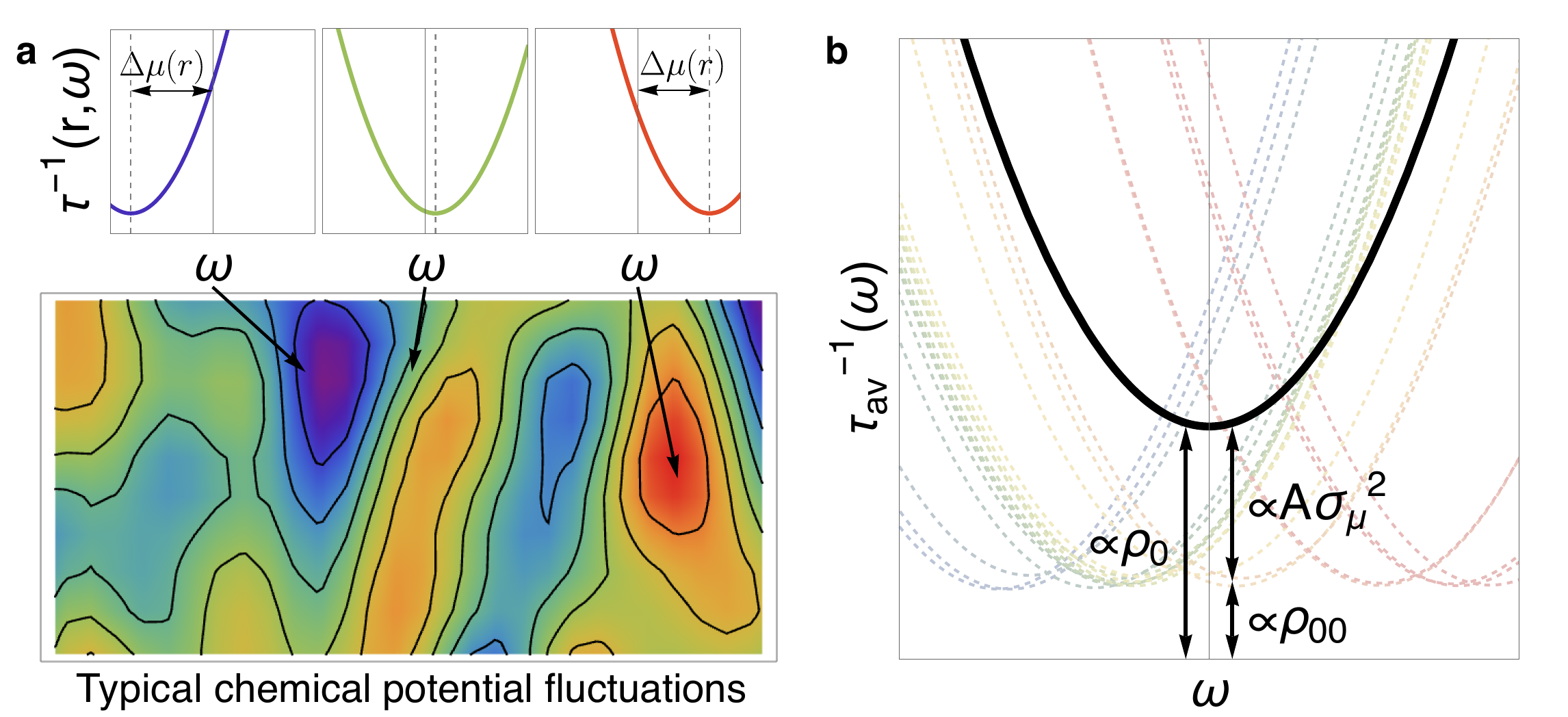}
 \caption{\textbf{Visualization of chemical potential fluctuations.} 
 {\bf a.} In a FL the electron scattering rate increases quadratically as a function of energy away from the Fermi energy, $\tau^{-1} \sim (\omega - \mu)^2$. 
Chemical potential fluctuations can locally shift the scattering rate $\tau^{-1}(\omega,r)$ depending on position $r$.
{\bf b.} Integrating over the local chemical potential fluctuations yields an average scattering rate $\tau^{-1}_{\rm av}$. The scattering rate at $\omega = 0$ is proportional to the residual resistivity $\rho_0$. In addition to the disorder scattering component $\rho_{00}$, the residual resistivity is increased by a term proportional to the chemical potential fluctuation variance $\sigma_\mu^2$ times the $A$ coefficient.
}
\label{chemical_potential_fluct}
\end{figure}

The above theory applies to correlated metals in the presence of conventional disorder scattering.
To account for the observed phenomenology at odds with standard FL theory, here we examine a different effect of disorder: in addition to impurity scattering, a metal can exhibit randomness in the form of large-scale fluctuations of the chemical potential $\mu(r)$. Such `charge puddles' have been observed in graphene samples~\cite{Martin2008, Deshpande2009, Zhang2009, Xue2011, Decker2011, Andrei_2012,Burson2013, Gibertini2012} on the $\mu$m scale; inhomogeneous charge transport in relation to disorder has been also discussed based on optical data of the compounds under study~\cite{Pustogow2021-Landau}. In strongly correlated organics close to a Mott-insulating state, where physics turns local, we expect length scales in the sub-micrometer range -- in any case much smaller than the sample size that exceeds 100~$\mu$m in all directions. 
If the fluctuations of $\mu(z)$ are smooth over distances where FL correlations are established, then the local scattering rate is of the FL form with a locally shifted chemical potential,
\begin{equation}
	\tau^{-1}_{\rm local}(r, \omega, T) = \tau^{-1}_{\rm FL}(\omega - \mu(r), T).
\end{equation}
This idea is sketched in Fig.~\ref{chemical_potential_fluct}a, where we show a typical plot of chemical potential fluctuations and how the local scattering time is shifted with the local $\mu(z)$. This effect arises because FL theory is applied locally, and the energy is determined by the highest fully occupied state — the local chemical potential — which can vary across space.

The scattering rate over larger distances $ \tau^{-1}_{\rm av} (\omega,T) = \int d\mu \; P(\mu) \; \tau^{-1}_{\rm local}(r, \omega, T)$ is obtained by averaging the local $\tau^{-1}(\omega,r)$ over the distribution $P(\mu)$ of chemical potential fluctuations, shown in Fig.~\ref{chemical_potential_fluct}b. The sole assumption concerning the distribution is that it is symmetric around $\omega = 0$. The average scattering rate becomes 
\begin{eqnarray}
	\tau^{-1}_{\rm av} (\omega,T) = \tau_{00}^{-1} + A' \sigma_\mu^2 + A' \left(\omega^2 + (\pi T)^2 \right) + \ldots,
\end{eqnarray}
where $\sigma_\mu^2$ is the variance of the chemical potential fluctuations.
Consequently, the residual resistivity obtains a contribution proportional to the $A$ coefficient,
\begin{equation}
	\rho_0 = \rho_{00} + \frac{3}{4\pi^2} A \sigma_\mu^2,
	\label{Eq:CentralRelation}
\end{equation}
where $\rho_{00}$ is the contribution from disorder scattering independent of correlation strength. Eq.~\eqref{Eq:CentralRelation} is the central result of this paper, showing that the residual resistivity in a correlated metal is linearly proportional to $A$, i.e. to the square of the effective mass enhancement $(m^{\star}/m)^2$, and to the variance of chemical potential fluctuations.

Note that chemical substitution, as employed in our experiments, increases disorder and therefore gives rise to stronger chemical potential fluctuations. Consequently, the variance $\sigma_\mu^2$ can be tuned via chemical substitution explaining the growth of $\frac{d\rho_0}{dA}$ in Fig.~\ref{STF-resistivity-all}d.

\begin{figure}
  \hspace{-0.05\textwidth}
 \includegraphics[width=0.45\textwidth]{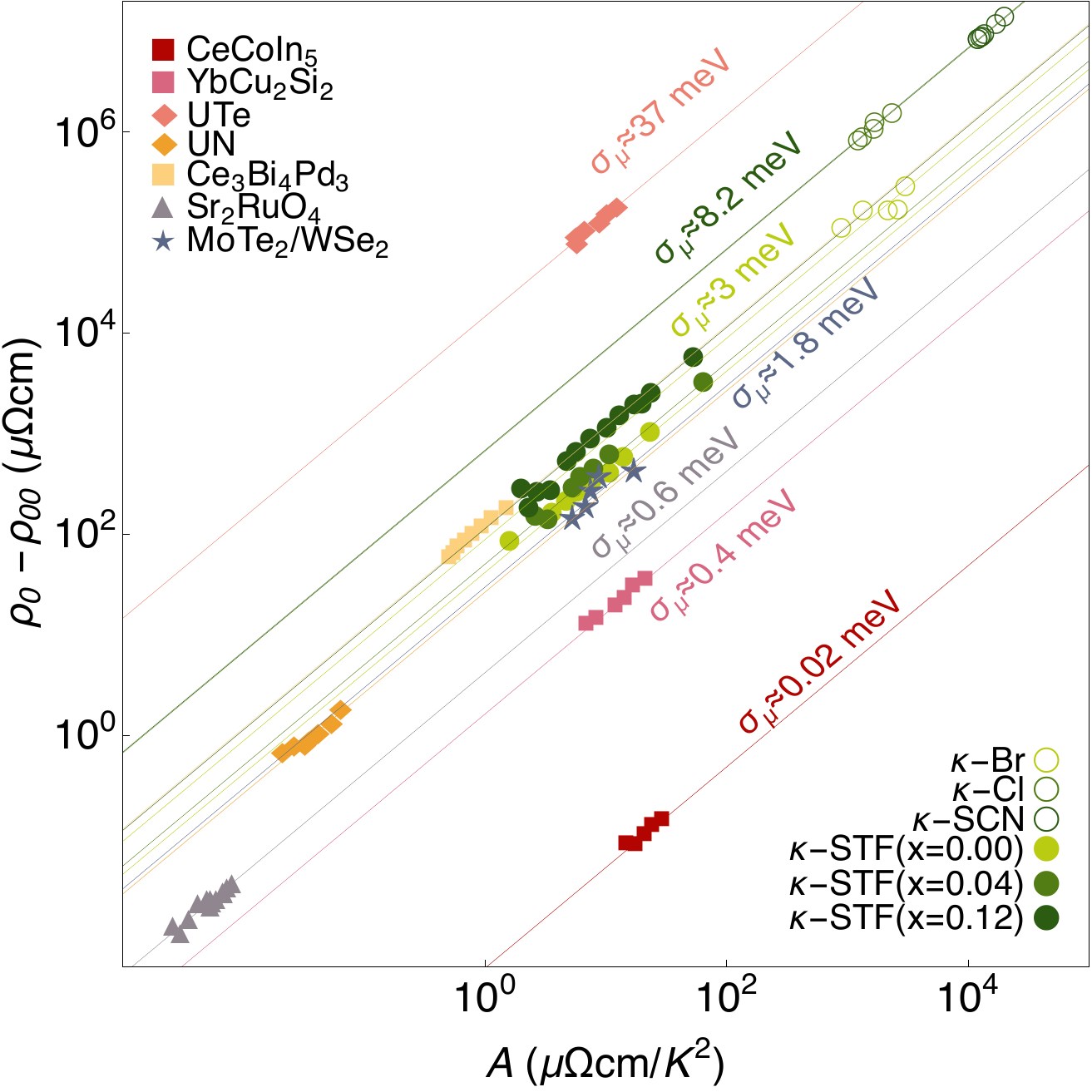}
 \caption{\textbf{Universal scaling of $A$-coefficient with residual resistivity.} 
The resistivity in a correlated Fermi liquid is given by Eq.~\ref{Eq-resistivity}. 
The linear relationship $\rho_0 \propto A\sigma^2_\mu$ from Eq.~\ref{Eq:CentralRelation} is observed for several organic \stf\ compounds (this work, see Fig.~\ref{STF-resistivity-all}d), other $\kappa$-phase organics, namely  $\kappa$-Cl, $\kappa$-SCN,  and $\kappa$-Br (see full chemical formulas in the footnote~\cite{kappa_organics}) ~\cite{Urai2019,Analytis2006,Strack2005}, \sro~\cite{Barber2018}, heavy-fermion systems~\cite{Ronning.2005,Alami-Yadri1998,Kushwaha2019,Link1992,Nakashima2003} and moiré MoTe$_2$/WSe$_2$~\cite{Zhao.2023}. 
The different solid lines on the log-log plot indicate a different degree of chemical potential fluctuations with variance $\sigma_\mu^2$. 
}

\label{scaling-rho0-A}
\end{figure}

{\em Universality -- } We find that the observed linear relation $\rho_0 \propto A$ is universal in a wide class of correlated metals, as can be seen by plotting $\rho_0-\rho_{00}$ versus $A$ in Fig.~\ref{scaling-rho0-A}.
In addition to the measured relation for \stf\ from the present work, we analyzed transport data in other organic compounds~\cite{Analytis2006,Strack2005,Urai2019}, \sro~\cite{Barber2018}, heavy-fermion systems~\cite{Ronning.2005,Alami-Yadri1998,Link1992,Nakashima2003,Kushwaha2019} and moiré MoTe$_2$/WSe$_2$~\cite{Zhao.2023}. For details on the data extraction, see the online Supplementary information.\cite{SupplInfo}

{\em Outlook -- } 
Our comprehensive transport results in organic \stf\ compounds reveal a relation between the residual resistivity and the $A$-coefficient in correlated metals that was previously unknown. We showed that this relation $\rho_0 \propto A$ can arise from chemical potential fluctuations. While we study here transport data only, it would be very interesting to compare the deduced chemical potential fluctuations $\sigma_\mu^2$ to direct measurements of such fluctuations. 
Scanning-tunneling microscopy (STM) provides a high-resolution method for probing local electronic properties, allowing to extract chemical potential variations at the nanoscale. In graphene, such fluctuations manifest as electron-hole puddles, which have been directly visualized in experiments on SiO$_2$ ~\cite{Deshpande2009, Zhang2009, Gibertini2012} and hexagonal boron nitride ($h$-BN) substrates~\cite{Xue2011, Decker2011}. Kelvin probe force microscopy (KPFM) has also been used to map the local potential landscape ~\cite{Burson2013}, offering an alternative approach which was applied with both SiO$_2$ and $h$-BN substrates. Additionally, original single-electron transistor measurements ~\cite{Martin2008} have provided insights into potential fluctuations by detecting changes in charge-carrier density with high sensitivity. Early theoretical models trying to explain this phenomenon suggested that charged impurities were the primary cause of inhomogeneities ~\cite{Rossi2008, Rossi2009, Adam2007}, while later works ~\cite{Gibertini2012} argued that intrinsic factors, such as corrugations in real materials, can also contribute. Moreover, similar puddles were found in insulating materials ~\cite{Borgwardt2016, Knispel2017, Bagchi2019}. 

Possible upcoming endeavors may include quantitative comparison of different types of randomness, e.g. site disorder upon partial chemical substitution within the conduction layer of the organics studied here and impurities within the anion layer introduced by x-ray irradiation~\cite{Sasaki2012,Furukawa2015,Urai2020,*Urai2019,Yamamoto2020}.
Comparing our transport-based estimates with these direct experimental techniques could help establish a more complete understanding of how disorder and interactions affect electrons in correlated materials. 

While it is natural to expect chemical potential fluctuations to appear in many metals, it is an open question whether our linear relation applies to all correlated metals. Specifically, the role of a `Fermi-liquid length scale' versus the length scale of the fluctuations warrants further theoretical analysis.
Such a study should elevate our current phenomenological theory -- with simple averaging of the scattering rate -- towards a microscopic calculation, taking also into account possible competing mechanisms. In particular, it is an important question whether the converse could also happen, namely that the $A$ coefficient would be directly affected by disorder.

On the experimental side, here we separately tuned disorder (through chemical substitution) and interaction strength (through physical pressure) in a controlled fashion. One might argue that pressure could also change the amount of disorder in a system. We note, however, that in \sro\ the behavior of $\rho_0$ is non-monotonic under pressure, yet the relation $\rho_0 \propto A$ is maintained throughout the van-Hove singularity~\cite{SupplInfo}. This suggests that indeed pressure separately tunes the effective correlation strength, and is not affecting the bare disorder potential.

Our results can be placed in a wider context of the ongoing effort to understand disordered interacting quantum materials. The here uncovered linear relation between residual resistivity and $A$-coefficient could serve as a new starting point for understanding how the Mott and other metal-insulator transitions behave in realistic, disordered materials.\\

\textbf{Methods:}

\footnotesize{
\textbf{Resistivity as a function of temperature:} The general form of the quasiparticle lifetime due to electron-electron interaction is quadratic in temperature and frequency, and can be written as follows
\begin{equation}
    \frac{1}{\tau(\omega,T)} = \frac{1}{\tau_{00}} + A' \left( \omega^2 + (\pi T)^2 \right) + \dots
    \label{lifetime_wo_chp}
\end{equation}

\noindent The relation that allows one to express conductivity through the quasiparticle lifetime is given by the Kubo formula\cite{Powell2009,Fetter1971,Kubo1957},
\begin{equation}
    \sigma = \frac{n e^2}{m} \int d\omega\, \left(-\dfrac{\partial f(\omega,T)}{\partial\omega}\right)\, \tau(\omega,T).
\end{equation}

\noindent The Sommerfeld expansion can be used to obtain the low-temperature, zero-frequency approximation
\begin{equation}
     \int d\omega\, \tau(T, \omega)\, f'(T, \omega) \approx \frac{\tau_{00}}{1 + \tau_{00} A' \pi^2 T^2 } - \frac{\pi^2 T^2}{6} \cdot 2 A' \tau_{00}^2 + \dots
\end{equation}\\
To extract the resistivity, we invert the expression for conductivity and expand it at low temperatures. This yields
\begin{equation}
    \rho = \frac{m}{n e^2} \left( \frac{1}{\tau_{00}} + \frac{4\pi^2}{3} A' T^2 + \dots \right),
\end{equation}

\noindent which is precisely in the desired form $\rho = \rho_{0} + A T^2$.

\textbf{Chemical potential fluctuations:} When chemical potential fluctuations are present, they introduce a correction to the frequency dependence of the scattering rate
\begin{equation}
    \frac{1}{\tau(\omega,T)} = \frac{1}{\tau_{00}} + A' \left( (\omega - \Delta\mu)^2 + (\pi T)^2 \right) + \dots
    \label{lifetime_w_chp}
\end{equation}

\noindent However, experimentally, these fluctuations are observed only as an average over the distribution, since the entire sample is measured
\begin{equation}
    \left. \frac{1}{\tau(T,\omega)} \right|_{\text{av}} = \int d(\Delta\mu)\, P(\Delta\mu)\, \frac{1}{\tau(T,\omega - \Delta\mu)}.
\end{equation}

\noindent Assuming the fluctuation distribution is symmetric around zero
\begin{equation}
    \left. \frac{1}{\tau(T,\omega)} \right|_{\text{av}} = \left( \frac{1}{\tau_{00}} + A' \sigma_\mu^2 \right)
    + A' \left( \omega^2 + (\pi T)^2 \right) + \ldots
\end{equation}

\noindent One can now calculate the resistivity using the method provided in the previous section
\begin{equation}
     \rho = \frac{m}{n e^2} \left[ \left( \frac{1}{\tau_{00}} + A' \sigma_\mu^2 \right) + \frac{4\pi^2}{3} A' T^2 + \ldots \right]
\end{equation}

\noindent From this, the linear relation between the residual resistivity and the temperature coefficient $A$ becomes evident.

\textbf{Experimental methods:}

As shown in Fig.~\ref{intuition}, the original BEDT–TTF molecule (left) has only sulfur atoms in its rings. In the modified version, BEDT–STF (right), two of these sulfurs are replaced by selenium atoms. Synthesizing samples with a mixture of both molecules enables us to tune the amount of local disorder, as it is proportional to the number of introduced BEDT-STF molecules, while keeping the overall structure the same.  

In this work \STF\ single crystals with typical dimensions of $1\times0.5\times0.1$~mm$^3$ were prepared by standard electrochemical oxidation \cite{Geiser1991,Saito2018,Saito2021,Saito2021a}. The pressure-dependent dc resistivity for $x=0.00$, $0.04$ and $0.12$ was measured within the $bc$-plane from room temperature down to $T=1.4$~K by the standard four-probe technique. Pressure was applied using a clamp piston pressure cell, as described in detail in Ref.~\cite{Saito2021a}.

Note that pressure affects the {\em bare} band mass, and following Drude theory this leads to a smooth change of the residual resistivity. This should be contrasted with the dramatic changes near a quantum critical point (QCP)~\cite{Lohneysen2007,Coleman2005}. In particular, the Mott metal-insulator transition (MIT)~\cite{Georges1996,Gebhard1997,Imada1998} features a diverging {\em effective} mass, as has been observed in various organic systems~\cite{Dressel2020} consistent with the present study.



}


\bibliography{Literatur}
\bibstyle{unsrt}

\acknowledgments
We thank Vladimir Dobrosavljevic, Simone Fratini, Sergio Ciuchi, Karsten Held, Christophe Berthod, Antoine Georges and Nicola Poccia for valuable discussions. The project was supported by the Deutsche Forschungsgemeinschaft (DFG).
A.E. and L.R. are supported by the Swiss National Science Foundation through Starting Grant No. TMSGI2\_211296.


\newpage
\onecolumngrid





\appendix

\section{Theory: Electron self-energy}

In this section, we consider the extended derivations which were presented in the "Methods" section of the main text. Here we consider the case where the chemical potential is fixed and introduce a general method that allows us to derive the resistivity through the quasiparticle lifetime.

The quasiparticle lifetime reflects the effects of interactions within the system, which influence the scattering rate and, consequently, the transport properties.

In the Fermi liquid model, because of the one-to-one correspondence between electrons and quasiparticles, Fermi-Dirac statistics apply to quasiparticles, and the Pauli exclusion principle remains valid. Under these assumptions, it can be shown that the imaginary part of the self-energy—and thus the quasiparticle scattering rate—follows a quadratic dependence on both energy and temperature as leading terms. Consequently, the general form of the quasiparticle lifetime takes the form

\begin{equation}
    \frac{1}{\tau(\omega,T)}= \frac{1}{\tau_{00}} +A' \left( \omega^2 + (\pi T)^2 \right) + \dots
    \label{lifetime_wo_chp}
\end{equation}

where $\frac{1}{\tau_{00}}$ part came from the disorder scattering processes. 

Conductivity can be calculated using the Kubo formula, where the dependence on the spectral function introduces the imaginary part of the self-energy and, therefore, the QP lifetime in the nominator.

\begin{equation}
    \sigma = \frac{n e^2}{m} \int d\omega f'(\omega,T)\tau(\omega,T).
\end{equation}

The integral looks similar to the integral used for the Sommerfeld expansion with the additional derivative. We can use the similar rule for the expansion

\begin{equation}
     \int d\omega\, \tau(T, \omega) f'(T, \omega) \approx \tau(T, 0) + \frac{\pi^2 T^2}{6} \frac{\partial^2 \tau(T, \omega)}{\partial \omega^2} \Big|_{\omega = 0} + \dots
\end{equation}\\


Substituting the expression for the imaginary part of the self-energy, we obtain the leading Sommerfeld expansion coefficients

\begin{equation}
    \quad \tau(T, 0) = \dfrac{\tau_{00}}{1+\tau_{00} A' \pi^2T^2 },
\end{equation}

\begin{equation}
     \tau''(0, 0) = -2 A' \tau_{00}^2.
\end{equation}

To extract the resistivity, we invert the conductivity expression and expand at low temperatures. This yields

\begin{equation}
    \, \rho = \frac{m}{n e^2} \left( \frac{1}{\tau_{00}} + \frac{4\pi^2}{3} A' T^2 + \dots \right)
\end{equation}

Compared with the standard temperature-dependent resistivity formula,\\ $\rho = \rho_{0} + AT^2$, one can easily get that: 

\begin{equation}
   \rho_{0} = \rho_{00}= \dfrac{m}{ne^2} \dfrac{1}{\tau_{00}},
\end{equation}

\begin{equation}
    A = \frac{m}{n e^2}\dfrac{4\pi^2}{3}A'.
\end{equation}

Here we introduced $\rho_{00}$, which specifies that we consider the non-fluctuating chemical potential. We will use this quantity for further comparison. 

Thus, we obtain the well-known quadratic temperature dependence of the resistivity, which arises naturally from the expansion of the QP lifetime  in the low-temperature regime.\\

\textbf{Chemical potential fluctuations}

Let us consider a nonhomogeneous chemical potential and assume that there are regions where it can be considered static and where locally FL theory works; however, over the sample it forms puddles, which have been already observed for some materials.

Let's set $P(\Delta \mu)$ as distribution of chemical potential fluctuations and then for the length scale greater than puddles' size we can find an average QP lifetime: 

\begin{equation}
    \frac{1}{\tau(T,\omega)}_{av} = \int d(\Delta\mu) P(\Delta \mu) \frac{1}{\tau(T,\omega-\Delta\mu)}.
\end{equation}

\begin{equation}
    \frac{1}{\tau(T,\omega)}_{av} = \frac{1}{ \tau_{00}} + A' (\pi T)^2 +A'\int d\Delta\mu P(\Delta \mu) (\omega-\Delta\mu)^2 + \dots
\end{equation}

Assuming the fluctuation distribution is symmetric around 0: 

\begin{equation}
    \frac{1}{\tau(T,\omega)}_{av} = \frac{1}{ \tau_{00}} + A' \left(\omega^2+(\pi T)^2\right) + A'\int d\Delta\mu P(\Delta \mu) (\Delta\mu)^2 + \dots
\end{equation}

\begin{equation}
    \frac{1}{\tau(T,\omega)}_{av} = \left( \frac{1}{\tau_{00}} + A' \sigma_\mu^2 \right)
    + A' (\omega^2 + (\pi T)^2 ) + \ldots
\end{equation}

One can see that the form of the expression is very similar from the one we get in previous section \ref{lifetime_wo_chp}. We can rewrite the "constant" term as follows.
\begin{eqnarray}
   \frac{1}{\tau_{0}} =  \left( \frac{1}{\tau_{00}} + A' \sigma_\mu^2 \right).
\end{eqnarray}

 One can calculate resistivity based on the method provided in previous section: 

\begin{equation}
     \rho = \frac{m}{ne^2} \left[ \frac{1}{\tau_0} + \frac{4\pi^2}{3} A' T^2 + \ldots \right] =\frac{m}{ne^2} \left[ \left( \frac{1}{\tau_{00}} + A' \sigma_\mu^2 \right) + \frac{4\pi^2}{3} A' T^2 + \ldots \right] 
\end{equation}

Let us compare it again with the main formula describing metallic behavior $\rho(T) = \rho_0 + AT^2$:

\begin{equation}
    \rho_0 = \frac{m}{ne^2} \left( \frac{1}{\tau_{00}} +  A' \sigma_\mu^2 \right) = \rho_{00} + \frac{m}{ne^2}A' \sigma_\mu^2,\qquad  A =  \frac{m}{ne^2}\frac{4\pi^2}{3} A'.
\end{equation}

Now one can easily express $\rho_0$ in terms of $A$ and see their linear dependence:
\begin{equation}
  \rho_0 = \rho_{00} + \frac{3}{4 \pi^2} A \sigma_\mu^2.
\end{equation}






\section{Experimental comparison}

In this section we analyse previously published resistivity data for a variety of metallic systems—including heavy‐fermion compounds, molecular (organic) metals, moiré heterostructures, and \(\mathrm{Sr}_2\mathrm{RuO}_4\).  Our goal is to extract the residual resistivity \(\rho_{0}\) and the quadratic‐temperature coefficient \(A\) in each case.  We process data via two methods:

\begin{enumerate}
  \item \textbf{Direct \(\rho(T)\) fitting.}  
    Whenever a single curve of \(\rho(T)\) was reported, we digitized the data and fitted it to
    \[
      \rho(T) \;=\; \rho_{0} \;+\; A\,T^{2}\,,
    \]
    thereby obtaining one \((\rho_{0},A)\) pair for each distinct condition (field, dose, pressure, …).

  \item \textbf{Indirect pairing from \(P\)-dependent plots.}  
    In studies where the authors presented \(A(P)\) and \(\rho_{0}(P)\) separately as functions of pressure \(P\), we digitized both curves and then paired the values of \(A\) and \(\rho_{0}\) at each common pressure.
\end{enumerate}


\subsection{Heavy fermions}

\begin{figure}[htbp]        %
  \centering                %
  \includegraphics[width=1\textwidth]{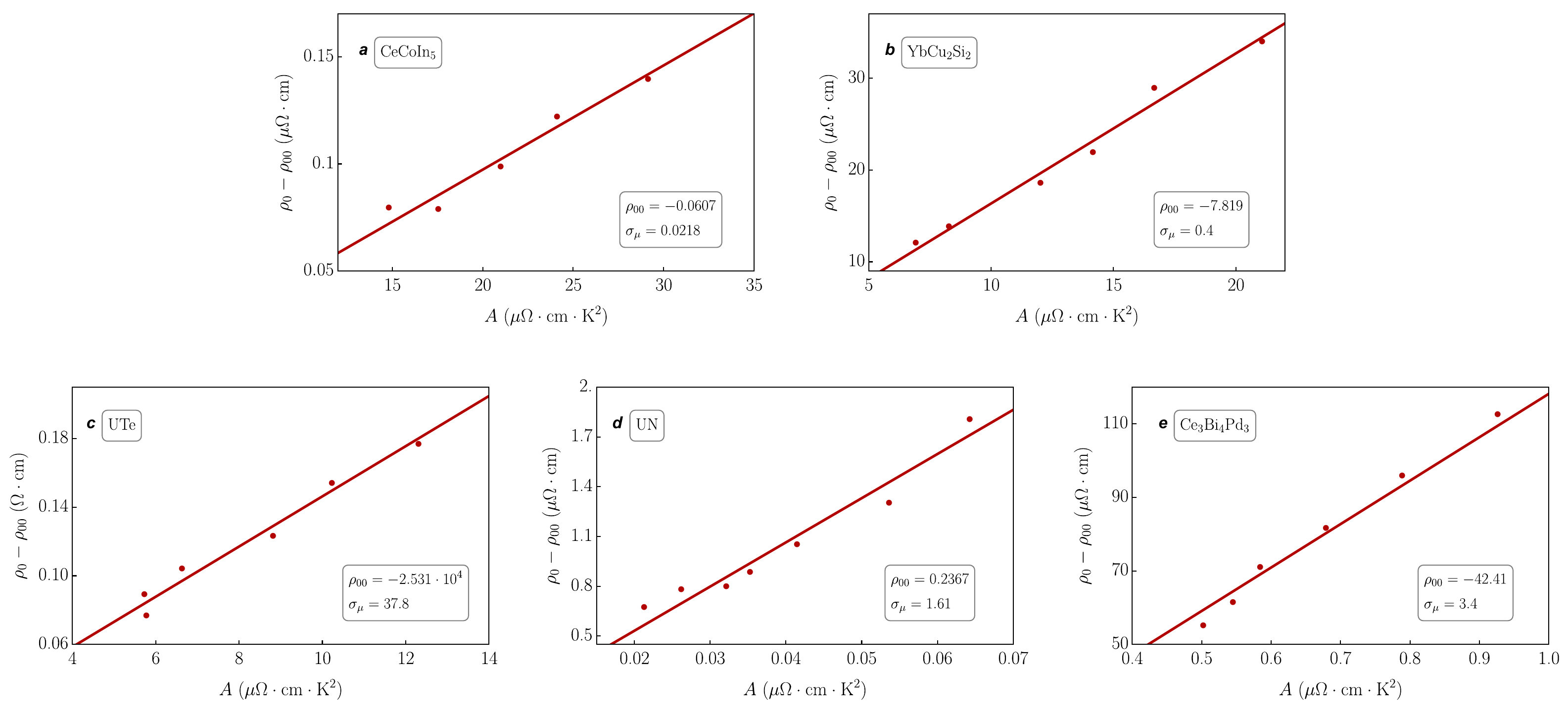}
  
  \caption[Heavy fermions]{\textbf{Temperature‐coefficient $A$ versus residual resistivity $\rho_0 - \rho_{00}$ for five heavy‐fermion compounds:} 
    (a) CeCoIn$_5$, (b) YbCu$_2$Si$_2$, (c) UTe, (d) UN, and (e) Ce$_3$Be$_4$Pd$_3$.  
    Solid symbols denote the measured data; solid lines are linear fits to $\rho_0 \;=\; \rho_{00} \;+\;\frac{3}{4\pi^2}\,\sigma_{\mu}^2\,A$,
    where the fit parameters $\rho_{00}$ and $\sigma_{\mu}^2$ (the latter expressed in meV) are listed for each material in its respective panel.
  }
  \label{fig:5heavyfermsuppl}  
\end{figure}

\paragraph{}  \textbf{CeCoIn$_5$}
 We digitized the $\rho$ vs.\ $T^2$ curves for CeCoIn$_5$ from Ronning \emph{et al.}\,\cite{Ronning.2005} ($Fig.$~$3(a)$ in that work, for $H=12.5$, $14$, $15$, $16.5$ and $18$ $T$) and applying the direct $\rho(T)$ fitting procedure described above, we extract the field-dependent pairs $(\rho_0,A)$.  These data appear in Fig.4 of the main text (log–log scale) and in Fig.~\ref{fig:5heavyfermsuppl}(a) of the Supplementary Information.

\paragraph{}  \textbf{YbCu$_2$Si$_2$}
Using the indirect pairing procedure described above, we digitized the pressure‐dependent curves $A(P)$ and $\rho_{0}(P)$ from Alami‐Yadri \emph{et al.}\,\cite{Alami-Yadri1998} ($Figs.~6$ and $7$ of that work).  To focus on the heavy‐fermion regime, we excluded points near the magnetic transition at $P\approx8.5$\,$GPa$ and retained only data for $P>8.5$\,$GPa$.  The extracted $(\rho_0,A)$ pairs exhibit a pronounced upturn and apparent divergence as $P$ approaches $\sim 12$$GPa$. These data appear in $Fig.4$ of the main text (log–log scale) and in $Fig.$~\ref{fig:5heavyfermsuppl}$(b)$ of the Supplementary Information.  

\paragraph{}  \textbf{UTe}
Applying the direct $\rho(T)$ fitting procedure described above, we digitized the \(\rho\) vs.\ \(T\) curves for UTe from Link \emph{et al.}\,\cite{Link1992} ($Fig.$~$1$ of that work, for $P=1.5$, $2.5$, $2.8$, $4.1$, $5.3$ and $6$ $GPa$), and obtain the \((\rho_0,A)\) pairs.  These points are plotted in $Fig.4$ of the main text (log-log scale) and in $Fig.$~\ref{fig:5heavyfermsuppl}$(c)$ of the Supplementary Information.

\paragraph{} \textbf{UN} 
Using the indirect pairing procedure described above, we digitized the pressure‐dependent curves $A(P)$ and $\rho_0(P)$ from Nakashima \emph{et al.}\,\cite{Nakashima2003} ($Figs.$~$5(a)$ and $5(b)$ of that work).  To remain within the Fermi‐liquid regime, only data for \(P\ge2\)\,GPa were retained $(P=2,\;2.5,\;3,\;3.5,\;4,\;5,\;7$ $GPa)$.  The resulting \((\rho_0,A)\) pairs are plotted in $Fig.$~$4$ of the main text (log-log scale) and in $Fig.$~\ref{fig:5heavyfermsuppl}$(d)$ of the Supplementary Information.

\paragraph{}  \textbf{Ce$_3$Bi$_4$Pd$_3$}
Applying the direct \(\rho(T)\) fitting procedure described above, we digitized the \(\rho\) versus \(T^2\) curves for Ce$_3$Bi$_4$Pd$_3$ from Kushwaha \emph{et al.}\,\cite{Kushwaha2019}. In $Fig.$~$5(b)$ of that work, the authors present $\rho(T)$ plotted as $\rho$ vs.\ $T^2$ for six magnetic fields $H=25$, $30$, $35$, $40$ ,$45$, $50$ and $59$ $T$.  The resulting field-dependent \((\rho_0,\,A)\) pairs are plotted in $Fig.$~$4$ of the main text (log–log scale) and in $Fig.$~\ref{fig:5heavyfermsuppl}$(e)$ of the Supplementary Information.

\subsection{\sro}

\begin{figure}[htbp]        %
  \centering                %
  \includegraphics[width=0.8\textwidth]{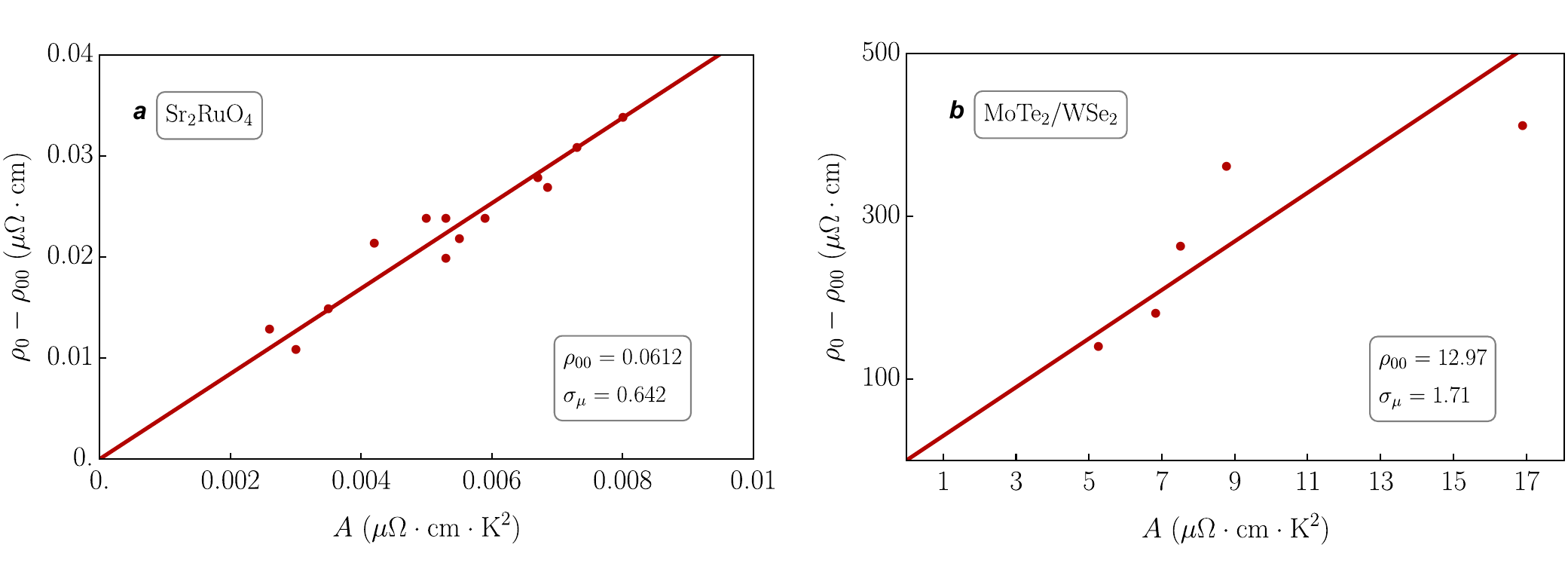}

  \caption[SrRu$_2$O$_4$ and Moire]{ \textbf{Temperature‐coefficient \(A\) versus residual resistivity \(\rho_0 - \rho_{00}\) for } 
    (a) SrRu$_2$O$_4$, (b) MoTe$_2$/WSe$_2$.  
    Solid symbols denote the measured data; solid lines are linear fits to $ \rho_0 \;=\; \rho_{00} \;+\;\frac{3}{4\pi^2}\,\sigma_{\mu}^2\,A$,
    where the fit parameters \(\rho_{00}\) and \(\sigma_{\mu}^2\) (the latter expressed in meV) are listed for each material in its respective panel.
  }
  \label{fig:SrMoire_suppl}  
\end{figure}

In order to check whether the observed phenomenology is universal to correlated metals or rather a specific feature of the Mott MIT, we evaluate transport data fully metallic systems that do not undergo a transition to an insulating state. 
We downloaded the uniaxial‐strain–dependent resistivity data for Sr$_2$RuO$_4$ from Barber \emph{et al.}\,\cite{Barber2018}, with SDW‐onset markers taken from Refs.~\cite{Grinenko2021,Li2022}. The strain changes from $0.00$ to $0.92$ $\%$.  The resulting strain‐dependent \((\rho_0,\,A)\) pairs are plotted in $Fig.$~$4$ of the main text (log–log scale) and in Fig.~\ref{fig:SrMoire_suppl}$(a)$ of the Supplementary Information.

\subsection{Moiré MoTe$_2$/WSe$_2$}

We included the resistivity data from Zhao  \emph{et al.}~\cite{Zhao.2023} on the moiré heterobilayer MoTe$_2$/WSe$_2$. We used the publicly available data from their Fig.~3a, at fillings $n = 1 + x$, where the tuning parameter is the electronic density $x$, where $x=0.15$, $0.20$, $0.25$, $0.30$, $0.35$. Since this is a 2d material, only the resistance $R_{\square}$ is reported. By fitting $R_{\square} = R_0 + A_{\square} T^2$ we still get a linear relation between $R_0$ and $A_{\square}$. For plotting purposes, in Fig.~3a of our main manuscript we converted the sheet resistance to a resistivity $\rho_0$ assuming a material thickness of 1nm. Note that the linear relation $R_0 \propto A_{\square}$ is unaffected by this plotting choice. The log-log plot is presented in the main text $Fig.$~$4$ and the linear scale plot can be found in Fig.~\ref{fig:SrMoire_suppl}$(b)$.

\subsection{Organics}

Beyond the $
\kappa\text{-}\bigl[(\mathrm{BEDT}\text{-}\mathrm{TTF})_{1-x}(\mathrm{BEDT}\text{-}\mathrm{STF})_{x}\bigr]_2\mathrm{Cu}_2(\mathrm{CN})_3$ compounds discussed in the main text, we also examined a broader set of $\kappa$-type organics to test the generality of our findings.

\paragraph{} \textbf{$\kappa\,$-(BEDT-TTF)$_2$Cu(SCN)$_2$}

We applied the direct $\rho(T)$ fitting procedure described above to the interlayer resistivity data of $\kappa\,$-(BEDT-TTF)$_2$Cu(SCN)$_2$ from Analytis \emph{et al.}\,\cite{Analytis2006} (Fig.~2 of that work).  After digitizing the low–$T$ portion of each curve, we extracted $(\rho_0,A)$ for the following X-ray irradiation doses:  
$
0,\;20,\;40,\;80,\;160,\;320\; \text{and},\; 470\;\mathrm{MGy}.
$
These points appear in Fig.~4 of the main text (log-log scale) and in Fig.~\ref{fig:kappanew_suppl}(a) of the Supplementary Information.

\paragraph{}\textbf{$\kappa\,$-(ET)$_2$Cu[N(CN)$_2$]Cl}

Applying the direct \(\rho(T)\) fitting procedure described above, we digitized the low–temperature \(\rho\) vs.\ \(T^2\) curves for \(\kappa\)-(ET)\(_2\)Cu[N(CN)\(_2\)]Cl from Urai \emph{et al.}\,\cite{Urai2019} (Figs.~3(c) of that work) at successive X-ray irradiation time \(t_{\rm irr}=70\) h under different pressures $P=19,\;26,\;32,\;41,\;\text{and}\; 59$ MPa. Obtained \((\rho_0,A)\) pairs plotted in Fig.~4 of the main text (log-log scale) and in Fig.~\ref{fig:kappanew_suppl}(b) of the Supplementary Information.  

\paragraph{}\textbf{$\kappa\,$-(BEDT-TTF)$_2$Cu[N(CN)$_2$]Br}  

Applying the direct \(\rho(T)\) fitting procedure described above, we digitized the low-temperature out-of-plane resistivity curves for both the high-resistance (HR) and low-resistance (LR) variants of \(\kappa\)-(BEDT-TTF)\(_2\)Cu[N(CN)\(_2\)]Br from Strack \emph{et al.}\,\cite{Strack2005} (Fig.~6(a)), measured under hydrostatic He-gas pressure for $P=0,\;170,\;350,\;1000,\;\text{and}\; 2000$ kbar. \((\rho_0,A)\) pairs for these pressures appear in Fig.~4 of the main text (log-log scale) and in Fig.~\ref{fig:kappanew_suppl}(c) of the Supplementary Information.

\begin{figure}[htbp]        %
  \centering                %
  \includegraphics[width=1\textwidth]{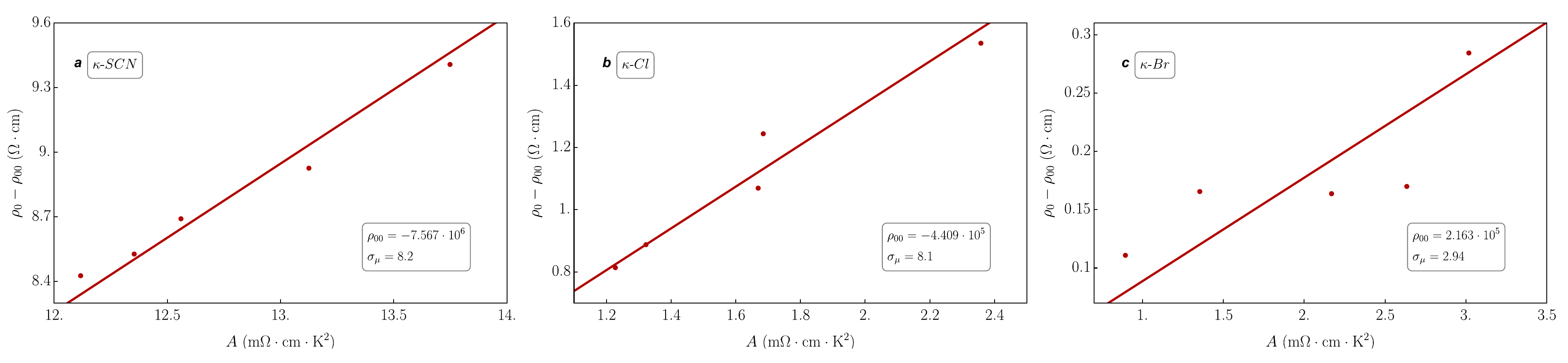}

  \caption[New organics]{\textbf{Temperature‐coefficient \(A\) versus residual resistivity \(\rho_0 - \rho_{00}\) for three $\kappa$-organics compounds:} 
    (a) $\kappa$-SCN, (b) $\kappa$-Cl, (c) $\kappa$-Br.  
    Solid symbols denote the measured data; solid lines are linear fits to $ \rho_0 \;=\; \rho_{00} \;+\;\frac{3}{4\pi^2}\,\sigma_{\mu}^2\,A$,
    where the fit parameters \(\rho_{00}\) and \(\sigma_{\mu}^2\) (the latter expressed in meV) are listed for each material in its respective panel.
  }
  \label{fig:kappanew_suppl}  
\end{figure}

For better comparison, we also added plots for the materials which were measured in the current work. Three sumples with different chemical substitution $x$ of BEDT-STF molecules were consider, where $x=0.00$, $0.04$, and $0.012$. The pressure changed from $4$ to $15$ $kbars$, the accurate range is indicated in $Fig.$~$2$(a)-(c) in the main text.

\begin{figure}[h!]        %
  \centering                %
  \includegraphics[width=1\textwidth]{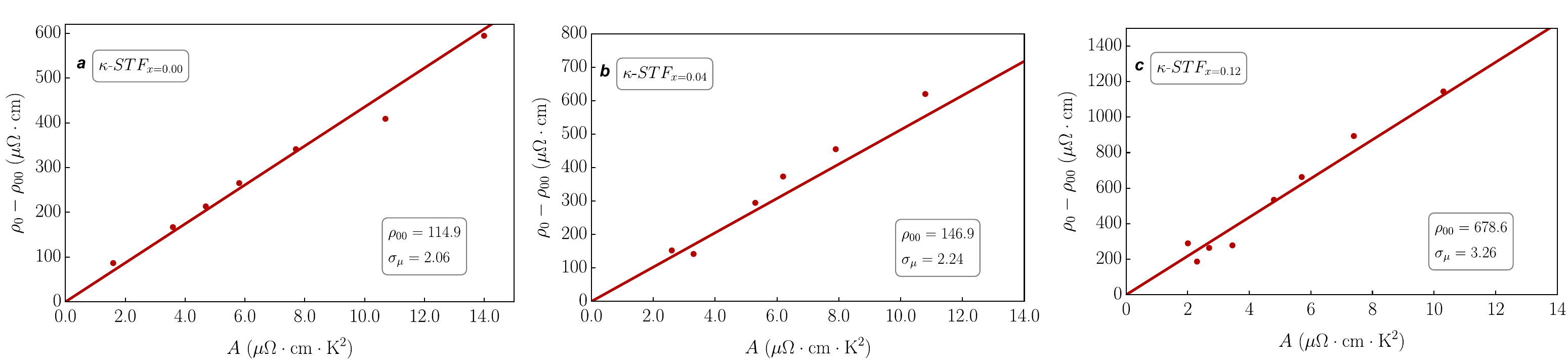}

  \caption[Our organics]{\textbf{Temperature‐coefficient $A$ versus residual resistivity $\rho_0 - \rho_{00}$  for three $\kappa$-$STF$-organics compounds which were measured in present work:} 
    (a) $\kappa$-$STF_{x=0.00}$, (b) $\kappa$-$STF_{x=0.04}$, (c) $\kappa$-$STF_{x=0.12}$.  
    Solid symbols denote the measured data; solid lines are linear fits to $ \rho_0 \;=\; \rho_{00} \;+\;\frac{3}{4\pi^2}\,\sigma_{\mu}^2\,A$,
    where the fit parameters \(\rho_{00}\) and \(\sigma_{\mu}^2\) (the latter expressed in meV) are listed for each material in its respective panel.
  }
  \label{fig:your-label}  
\end{figure}




\end{document}